\documentclass[twocolumn, pre, showpacs]{revtex4-1}

\usepackage[normalsize,tight]{subfigure} \usepackage{amssymb}
\usepackage{amsmath} \usepackage{graphicx}
\usepackage[pdfpagelabels=false]{hyperref}
\usepackage{cleveref} 

\begin{document}

\def\halfrs{\vspace*{-3ex}} \def\rs{}
\newcommand{\xhdr}[1]{\paragraph*{\bf #1}} \newcommand{\alg}{GCE}
\newcommand{\etal}{et al.}  \newcommand{\apriori}{\textit{a priori}}
\newcommand{\ie}{i.e.}  \newcommand{\etc}{etc.}
\newcommand{\eg}{e.g.}  \newcommand{\Fscore}{F${}_1$-score }
\newcommand{\Fscores}{F${}_1$-scores }

\title{Seeding for pervasively overlapping communities}

\author{Conrad Lee} \email[Contact at ]{conradlee@gmail.com}
\author{Fergal Reid} \author{Aaron McDaid} \author{Neil Hurley}
\affiliation{Clique Research Cluster, UCD CASL, 8 Belfield Office
  Park, Clonskeagh, Dublin 4, Ireland}

\date{\today}

\begin{abstract}
  In some social and biological networks, the majority of nodes belong
  to multiple communities. It has recently been shown that a number of
  the algorithms that are designed to detect overlapping communities
  do not perform well in such highly overlapping settings. Here, we
  consider one class of these algorithms, those which optimize a local
  fitness measure, typically by using a greedy heuristic to expand a
  seed into a community. We perform synthetic benchmarks which
  indicate that an appropriate seeding strategy becomes increasingly
  important as the extent of community overlap increases. We find that
  distinct cliques provide the best seeds.  We find further support
  for this seeding strategy with benchmarks on a Facebook network and
  the yeast interactome.
\end{abstract}


\maketitle

\section{Introduction}
\label{introduction}

In social networks such as Facebook, users commonly belong to multiple
communities that may correspond to family, school, and professional
groups \citep{marlow-2009}. Similarly, in biological networks some
proteins belong to multiple functional complexes
\citep{sawardecker-2009, palla-2005}. Driven by these observations,
many community assignment algorithms have been developed that are, in
principle, capable of detecting overlapping communities.

\citet{Ahn2010} have recently pointed out that the extent of this
overlap is often greater than previously assumed, to the point where
the community structure may be described as \textit{pervasively
  overlapping}, a condition in which nearly all nodes belong to
multiple communities. As noted by \citeauthor{Ahn2010}, pervasive
overlap violates one of the only commonly accepted properties of
network communities, namely, that they contain more internal than
external edges \citep{radicchi-2004,newman-2004-2,lancichinetti-2009,
  clauset-2005}. As a result, some of the basic assumptions behind
many of these overlapping community assignment algorithms are
inappropriate.

In \citet{Lee2010}, we benchmarked several of these algorithms and
found that although they are designed to detect overlapping
communities, they perform poorly on networks with pervasively
overlapping communities. Our purpose here is to discover why a certain
class of these algorithms perform poorly on such networks: those which
optimize a local fitness measure, typically by using a greedy
heuristic to expand a seed node or edge into a community
\citep[including ][]{lancichinetti-2009, clauset-2005, baumes-2005,
  Bagrow2008, Mislove2010, Luo2008, Lee2010, Havemann2010}. By
expanding a set of seeds independently, either in serial or in
parallel, the objective is to discover all communities in the network.

Using synthetic data, we demonstrate that in networks with pervasively
overlapping communities, an appropriate seeding strategy is crucial,
and that several proposed seeding strategies perform poorly. Our
results, which are supported by benchmarks on empirical networks,
indicate that the best strategy is to seed with distinct cliques.

\section{Background}
\label{method}
In this section, we describe the core concepts and terminology of the
local algorithms addressed by this paper: community fitness functions,
greedy expansion, and community distance measures. We also describe
some of the key points on which these algorithms differ. We neither
fully specify these algorithms nor exhaustively list the differences
between them---here we merely provide a basic understanding so that
readers may see how all of these methods can benefit from an improved
seeding strategy.

We begin with terminology. Given an undirected, unweighted graph
$G(V,E)$, we define a community $C$ to be a set of vertices in
$V$. Two sets of edges are associated with each community:
$E_{in}(C)$, the edges induced by $C$, and $E_{out}(C)$, the set of
all edges that have one end in $C$ and the other end in the complement
of $C$. We will find it useful to refer to the \textit{frontier} of
$C$, which we define as the set of all nodes in $V$ that are adjacent
to, but not included in, $C$. In what follows, we assume the graph is
connected and unweighted, and simple, although it is trivial to extend
the following methods to the case of weighted or disconnected graphs.

A community fitness function $F(C)$ assigns a real value to a
community $C$, where a community with a higher fitness value
corresponds to a better defined community.  For example, the fitness
function used in the experiments of \citet{baumes-2005} is defined as
\begin{equation}
  F(C) = \dfrac{|E_{in}(C)|}{|E_{in}(C)| + |E_{out}(C)|},
  \label{baumes-fitness}
\end{equation}
which returns values between 0 and 1, with a trivial maximum when
$C=V$.

It will also be useful to be able to quantify the similarity of two
communities. Along the lines of \citet{baumes-2005}, we define a
community distance measure. We choose a symmetric measure of community
distance that can be thought of as the \textit{percent
  non-embedded}. Given two communities $C$ and $C'$, this measure is
defined as
\begin{equation}
  \delta(C,C')= 1 - \dfrac{|C \cap C'|}{ \min (|C|,|C'|)},
  \label{percent-non-embedded}
\end{equation}
which can be interpreted as the proportion of the smaller community's
nodes that are not embedded in the larger community. We will refer to
a community (or seed) that is within some distance $\epsilon$ of some
other community (or seed) of equal or larger size as a
\textit{near-duplicate} of that community.

We now move on to the idea of greedy expansion. Given such a fitness
function and a seed (which is typically a node or an edge) the seed
will is expanded into a community, which we will refer to as the
seed's target community. We begin by initializing $C$ to be the
seed. We then expand $C$ step-wise, one node at a time. At each
expansion step, we decide which node to add to $C$ by looping over
each node $v$ in the frontier of $C$, and calculating how the fitness
of $C$ would increase if $v$ were added to it. If no $v$ would
increase the fitness of $C$, then we terminate the expansion and
return $C$. Otherwise, we add to $C$ the $v $ that would most increase
the fitness of $C$, and proceed to the next expansion step.

While the fundamental strategy of many algorithms closely resembles
the one that we have just presented, they differ on their seeding
techniques (discussed in \cref{seeding}), the details of their
expansion (some are not strictly greedy), their fitness functions, and
their stopping criteria.  For example, rather than adding only one
node at a time, \citet{Luo2008}'s algorithm adds several nodes at each
expansion step. Another variant on the greedy expansion is to
alternate between expansion steps and contraction steps. The
algorithms of \citet{lancichinetti-2009} and \citet{Luo2008} include
such contraction steps, during which nodes whose removal from $C$
would result in increased fitness are purged from $C$.

Additional differences between these fitness functions are largely due
to a particular problem: the stopping criterion. As pointed out by
others \citep{lancichinetti-2009,Mislove2010}, \cref{baumes-fitness}
has the problem that as a community expands, the fitness function may
never reach a local maximum until the entire graph is included in one
big community. In other words, the local maximum of fitness does not
serve as a good termination criterion.

In the present work we benchmark the performance of four of these
expansion strategies, testing each with five seeding strategies. Our
purpose here is to discover how the seeding strategies are affected by
highly overlapping community structure. Because the focus here is on
seeding strategies, we describe these in the next section in
detail. Due to space constraints we will only briefly list the four
expansion techniques benchmarked below are:
\begin{itemize}
\item \textit{Greedy Clique Expansion (GCE)} as in \citet{Lee2010},
  which essentially greedily optimizes the fitness function in
  \cref{baumes-fitness} with the addition of a resolution parameter,
  $\alpha,$ which we leave set at its default value of 1.0 for the
  experiments below. This technique is based on and very similar to
  \citet{lancichinetti-2009},
\item \textit{Local Modularity} as in \citet{clauset-2005},
\item \textit{Normalized Conductance}, as in \citet{Mislove2010}, and
\item \textit{Iterative Scan}, which also uses the fitness function in
  \cref{baumes-fitness}, although expands using ``scans'' rather than
  in a strictly greedy fashion.
\end{itemize}
We refer the reader to the original papers in which these methods were
introduced for details.

\section{Seeding techniques}
\label{seeding}
Much analysis of local methods neglects the problem of seeding
altogether, treating seeds as user-provided parameters
\citep{clauset-2005,Bagrow2008, Luo2008}. Other literature suggests
seeding techniques that work for only a narrow range of situations,
such as using search engine results for seeding communities of web
pages \citep{Andersen2006}, or profile attributes
\citep{Mislove2010}. However, it is often not possible to obtain such
data, or such seeding may lead to communities of an undesired
scale---for example, seeding with all nodes of a particular gender,
dorm, or academic major may produce seeds that are larger than
desired. We now list schemes that discover seeds from graph topology
alone, and which we have implemented for our experiments in
\cref{verification}.  We note that all of the following techniques
have been proposed specifically for detecting overlapping communities.

\begin{itemize}
\item \textit{Random Seeding} \cite{lancichinetti-2009}.  Take as a
  seed a random node from the graph that has not yet been assigned to
  any community.  Expand that seed until that seed's community has
  been found (by greedily optimizing the fitness function in
  \cref{baumes-fitness} to its first local maximum). Repeat this
  process until all nodes have been assigned to at least one
  community.
\item \textit{Rank Removal} (\textit{RaRe} and \textit{RaRe Core})
  \cite{baumes-2005}. Nodes are first ranked by some measure of
  importance; the authors suggest PageRank works best. The highest
  ranked nodes are then iteratively removed from the graph, and the
  graph begins to break into disconnected components. When the size of
  the largest component is smaller than some user-defined value (which
  we will call the ``maximum core size''), one is left with
  cores. Cores below some minimum size are removed. Once the cores
  have been found, each of the removed vertices is added to any core
  to which it is connected. In our implementation we used all
  parameter settings as suggested in \citet{baumes-2005} (including
  PageRank for ranking the nodes, the maximum core size set to
  fifteen, and the minimum core size to three). We also implemented a
  version where the removed nodes are not added back to the cores---we
  refer to this strategy as \textit{RaReCore}.
\item \textit{Link Aggregate (LinkAgg)} \cite{baumes-2005:2}. Like in
  RaRe, nodes are first ranked, typically using PageRank. Next a list
  of seeds is created, which is initially empty. In order of
  decreasing rank, each node is then assigned to every seed whose
  fitness it would increase. If a node is not assigned to any seed, it
  becomes a new seed and is appended on to the list of seeds.
\item \textit{Distinct Cliques} \cite{Lee2010} Distinct cliques are
  the seeds used by the GCE algorithm described above. They are
  maximal cliques that are not within a given distance $\epsilon$ of
  any other maximal cliques of equal or greater size, where distance
  is as in \cref{percent-non-embedded}. That is, they are a set of the
  maximal cliques for which no near-duplicates exist. This seeding
  technique has two parameters, $k$, which specifies the minimum
  clique size, and $\epsilon$, which determines how similar to maximal
  cliques must be for them to be considered near duplicates. In our
  benchmarks, we set $k=4$ and $\epsilon=0.35$ unless otherwise
  noted. Furthermore, to speed up the removal of near duplicates, we
  use the CCH heuristic as suggested in \cite{Lee2010}.  This method
  is labeled as ``DistinctC'' in the figures.
\item \textit{Clique percolation}. \citet{palla-2005} introduced the
  concept of clique percolation to find overlapping communities, and
  the method has become well known. In his comprehensive review
  article, \citet{fortunato-2010} notes that the method has problems
  with leaf-nodes which are part of a community but not part of a
  large enough clique: ``Another big problem is that on real networks
  there is a considerable fraction of vertices that are left out of
  the communities, like leaves. One could think of some
  post-processing procedure to include them in the communities, but
  for that it is necessary to introduce a new criterion, outside the
  framework that inspired the method.'' The greedy expansion framework
  that we use to expand seeds here corresponds to the type of
  post-processing procedure suggested by
  \citeauthor{fortunato-2010}. Thus we test this suggestion, using
  k-clique communities as seeds. This seeding strategy includes one
  parameter, $k$, which we set to four in the following benchmarks
  unless otherwise noted.
\end{itemize}

In all of the experiments below, after the seeds have been expanded we
perform a post-processing step that removes near-duplicate
communities. We compare all pairs of communities, and if a smaller
community is a near duplicate of a larger community (as defined above,
with $\epsilon=0.4$), we remove it. If two equally-sized communities
are near-duplicates, we randomly remove one of them.  For this reason,
even when expansion techniques are seeded with the same method, they
can detect differing numbers of communities, as we will see in the
following section. Some expansion techniques expand several of its
seeds into near duplicate communities, which are filtered out, while
others expand them into distinct communities..

\section{Experiment methodology and results}
\label{verification}
We now evaluate the performance of the seeding strategies described in
\cref{seeding}. We begin by presenting the result of experiments on
synthetic graphs, and then proceed to results on empirical data.

\subsection{Synthetic Benchmarks}
An overview of our synthetic benchmark procedure is as follows: first,
we create a synthetic graph which, by construction, contains
communities planted in it. We will refer to these communities as the
\textit{ground truth} communities. Next, we run the algorithm on this
graph and call the returned communities the \textit{found}
communities. Finally, we use some metric to compare the similarity of
the ground truth communities to the found communities.
\begin{figure*}[t]
  \subfigure[]{
    \includegraphics[scale=1.0]{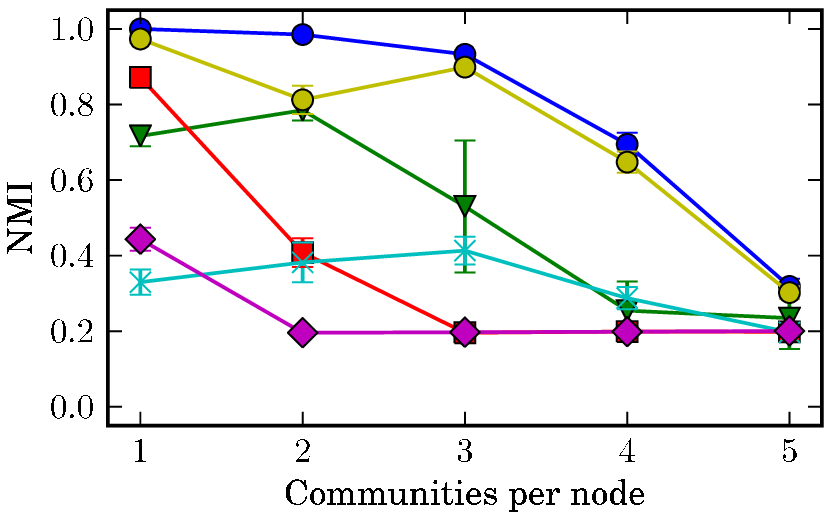}
    \label{gce_synthetic}
  } \subfigure[]{
    \includegraphics[scale=1.0]{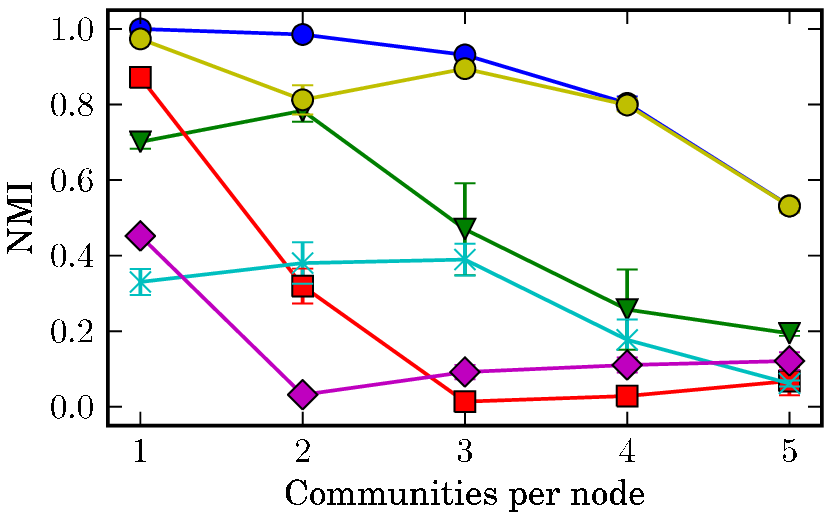}
    \label{localmod_synthetic}
  } \subfigure[]{
    \includegraphics[scale=1.0]{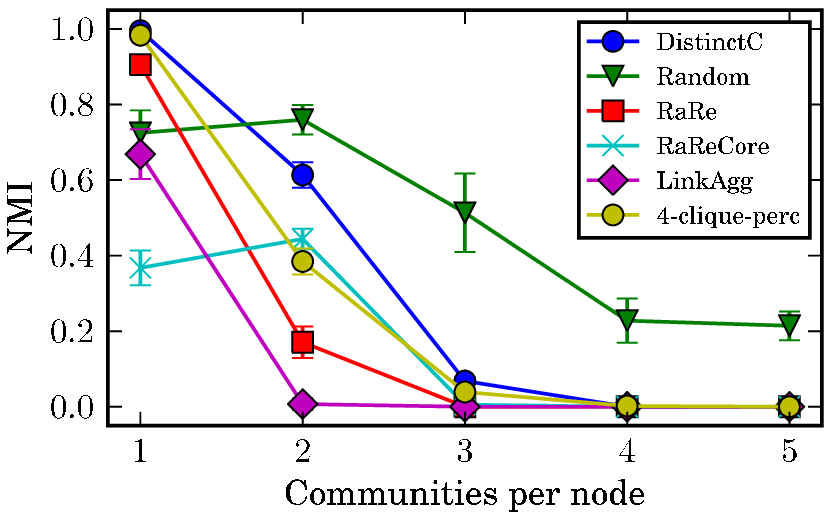}
    \label{iterscan_synthetic}
  } \subfigure[]{
    \includegraphics[scale=1.0]{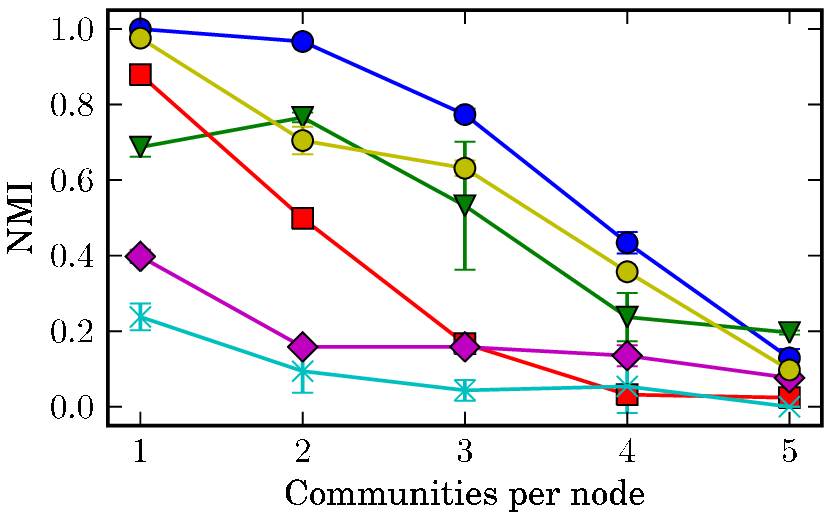}
    \label{normconduct_synthetic}
  }

  \caption{\label{synthetic_results} Each figure uses a different
    expansion strategy: (a), (b), (c), and (d) correspond to GCE, Local
    Modularity, Iterative Scan, and Normalized Conductance,
    respectively. Points correspond to the performance of various seeding
    strategies when run on synthetic graphs. The amount of community
    overlap in these graphs increases along the x-axis. Each point is
    the average of 10 runs, error bars indicate the standard deviation.}
  \vspace{-4mm}
\end{figure*}

To construct synthetic graphs, we use the \textit{LFR} specification
of \citet{lancichinetti-2009:2}, which allows one to create graphs
with realistic properties such as scale-free degree and community size
distributions. Our purpose here is to evaluate how various seeding
strategies perform on graphs with pervasively overlapping community
structure. As such, we here specify a series of five LFR graphs which
each contain 2000 nodes, but with increasing amounts of community
overlap. In the first graph, communities are disjoint; in the second
graph, each node belongs two communities, and so on, until in the
fifth graph each node belongs to five communities. To allow each node
to belong to an increasing number of communities, the degree of each
node increases in each successive graph specification. In the first
graph, the average degree is 18, in the second graph, the average
degree doubles to 36, and so on, until in the fifth graph the average
degree is 90.  The other parameters remain the same over all five
graph specifications: $N=2000$, $k_{max}=120$, $\mu=0.2$, $\tau_1=2$,
$\tau_2=2$, $C_{min}=60$, and $C_{max}=100$.

To measure the similarity of ground truth communities and found
communities, we use normalized mutual information (NMI), an
information-theoretic similarity measure.  This measure is normalized
such that the NMI of two sets of communities is 1 if they are
identical, and 0 if they are totally independent of each
other. \citet{danon-2005} first applied NMI to the problem of
evaluating the similarity of two sets of communities, but defined the
measure only for partitions. In our benchmarks, we employ a variant of
NMI introduced by \citet{lancichinetti-2009} that is defined for
covers, in which nodes may belong to multiple communities\footnote{For
  creating the LFR graphs and measuring overlapping NMI, we use the
  implementations provided by the authors, both of which are freely
  available at
  \url{http://sites.google.com/site/andrealancichinetti/software}.}.

The results in \cref{synthetic_results} show that the distinct clique
seeding strategy and the percolated-clique seeding strategy perform
best in this benchmark. They can provide good results even when every
node belongs to three or four communities. RaReCore and Link Aggregate
are wholly unsuited to this kind of network, even when the community
structure has no overlap.

Significantly, for three of the expansion methods, we see that
performance depends more on the seeding strategy than on the expansion
method. In contrast to the other three expansion methods, the
clique-based seeding methods to not work well with Iterative Scan.  We
discuss the possible causes of these results in \cref{discussion}.

\subsection{Empirical benchmarks}
Ideally, the empirical datasets would each come with a complete ground
truth, just like the synthetic datasets. However, the empirical
networks for which complete ground-truth data is known (such as
Zachary's Karate Club), are small networks that tend to have community
structure which is easy to recover. We were able to find two larger
networks that contain incomplete ground truths: Caltech's Facebook
network and the yeast interactome.

\begin{figure*}[t]
  \centering
  \includegraphics[]{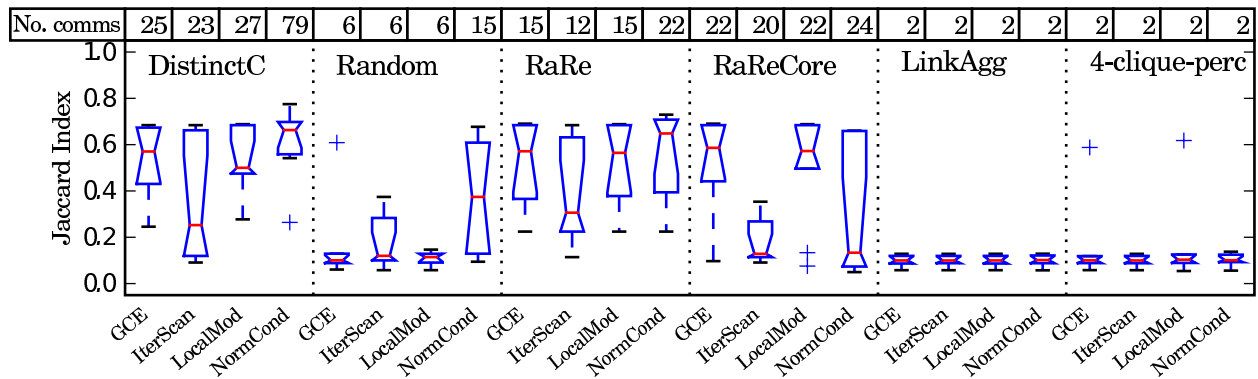}
  \label{caltech_boxplots}
  \caption{These boxplots indicate the ability of each seeding
    strategy to recover communities that correspond to the residential
    houses when run on Caltech's Facebook network. Whiskers represent
    lowest and highest datum still within 1.5 IQR of the lowest and highest
    quartiles, respectivley.}
  \vspace{-1mm}
\end{figure*}
\begin{figure*}
  \centering
  \includegraphics[]{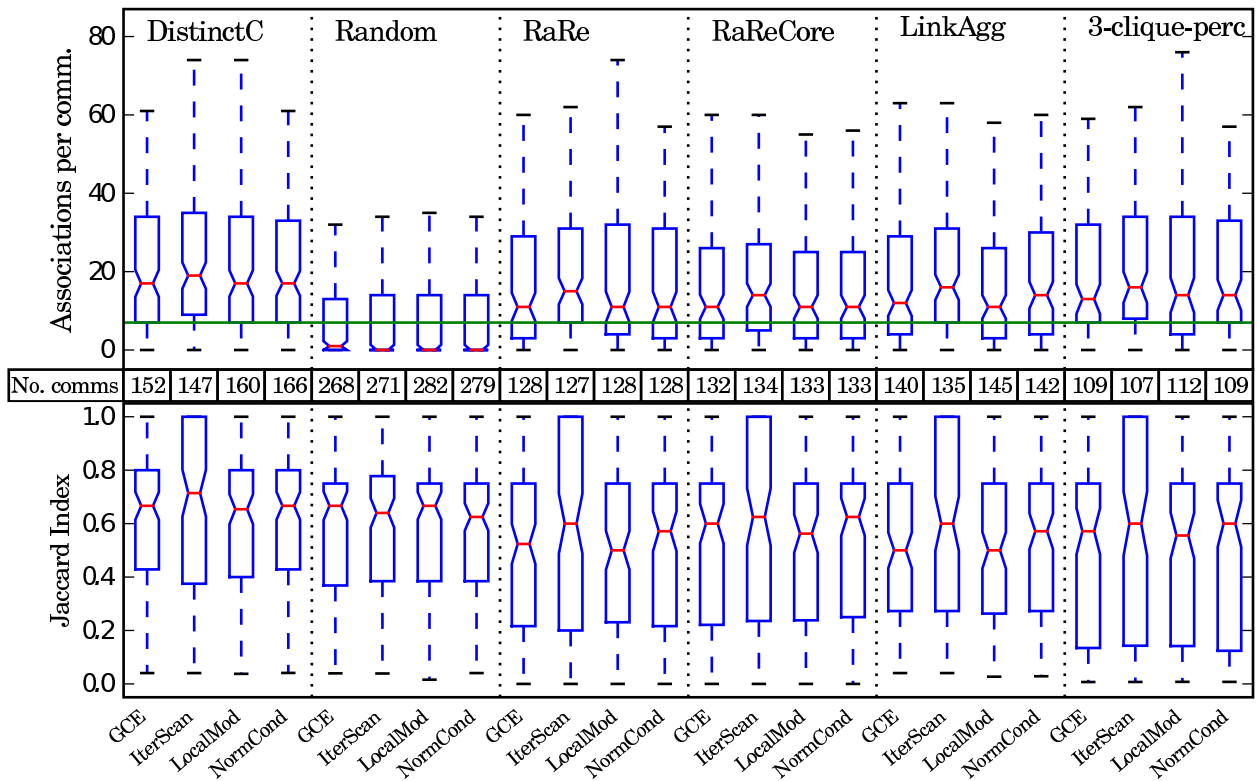}
  \caption{Boxplots indicating the ability of each seeding strategy to
    recover protein complexes from the CYC dataset when run on the
    yeast interactome. The lower row of boxplots was created by taking
    the Jaccard index of the most similar found community for each
    protein complex. The upper row of indicates how many times each
    found community was significantly associated with a function in
    the Gene Ontology.}
  \label{yeast_boxplots}
\end{figure*}

The Caltech data set consists of 769 nodes and 16656 edges. The data
originates from a Facebook employee, and includes all Caltech Facebook
users from June 2005 (excluding isolates) and the friendships between
them. Seventy-eight percent of these users indicated to which
residential house they belong. The paper in which this data set
appeared, \cite{traud-2009}, explains the significance of the
residential housing system:
\begin{quote}The undergraduate ``House'' system at Caltech, appearing
  in lieu of dormitory residence in our data, is modeled after the
  Oxbridge college system. Caltech's Housing system impacts student
  life enormously, both socially and academically, and is even used by
  the university as one of its primary selling points in attracting
  new undergraduates. At the beginning of their first year at Caltech,
  undergraduate students choose one of the eight Houses and usually
  remain a member of it throughout their collegiate career.
\end{quote}
Given this description, we expect any good community assignment
algorithm to detect the residential houses as communities.

However, in addition to the houses, there may exist additional
communities that correspond to, for example, study groups or sports
teams. The collection of houses is therefore an incomplete ground
truth, and a reasonable benchmark should not penalize an algorithm for
detecting additional communities. The consequence is that while we are
able to measure the sensitivity of the algorithms with respect to the
partial ground truth, we cannot measure the specificity of the methods
(i.e., we cannot distinguish false positives from true positives.)
For this reason, we abandon NMI as our metric of similarity between
the ground truth and found communities, and instead measure
sensitivity to the ground truth communities by matching each ground
truth community with its most similar counterpart in the set of found
communities.

Specifically, our benchmarking procedure is as follows. We first run
the community finding algorithm and are left with a set of found
communities. We remove all nodes for which house data is unknown
(22\%) from both the found communities and the ground truth. For each
house in the ground truth, we find the most similar found community,
where similarity is measured using the Jaccard index.  For each
algorithm, we keep track of these maximum similarities and present
their distribution in a boxplot in \cref{caltech_boxplots}.  Although
this evaluation does not return a number indicating the specificity of
each method, we note that none of the algorithms ``cheated'' this
shortcoming of the benchmark by simply finding a huge number of
communities.

In the bottom row of boxplots in \cref{yeast_boxplots}, we display the
results of a similar benchmark based on data from the yeast
interactome. To construct this protein-protein interaction network, we
used the Combined-AP/MS data, which contains 9070 interactions between
1622 proteins \cite{CombinedAPMS}. The protein complexes identified by
\citet{pu2009} in the CYC dataset provide an incomplete ground truth
for this network. Both the yeast interactome and the protein complex
dataset are publicly available \footnote{Combined-AP/MS data available
  at \url{http://interactome.dfci.harvard.edu/S_cerevisiae/}; CYC
  protein complex dataset available at
  \url{http://wodaklab.org/cyc2008/}}. This ground truth is peculiar
in that it contains a large number of communities which contain either
two or three nodes or that contain no edges whatsoever in the
interactome data. The latter condition may be a result of the
incompleteness of the protein-protein interaction network, which is a
well-known problem. We decided that for any community detection
algorithm to have a reasonable chance of detecting a protein complex
based on the network alone, it must contain at least one
triangle. After filtering out all ground truth complexes except those
with at least one triangle, 126 remained. Because many protein
complexes are small (and this would presumably be known by anyone
using community finding algorithms to detect protein complexes), we
set $k=3$ for both the distinct clique and clique percolation seeding
schemes for this benchmark.

Unlike the Caltech dataset, the yeast interactome has been the focus
of much research, allowing us to run a second benchmark which sheds
some light on specificity.  We use version 2.0 of FuncAssociate
(introduced in \cite{Berriz2009}) to test whether the set of proteins
that make up each community is more enriched in any Gene Ontology (GO)
attributes than one would expect by chance. FuncAssociate corrects for
multiple hypothesis testing using empirical resampling, and takes into
account that our experiment selected from only those proteins
contained in the Combined-AP/MS interactome.\footnote{Our purpose here
  is, given a community, to decide whether that community is
  significantly associated with any of several GO attributes. Because
  for each community we test several hypotheses, we have to correct
  for multiple hypothesis testing. If instead we were given a
  \textit{set} of communities, and our goal were to say which
  communities in that set are associated with any of the several GO
  attributes, then we would also want to further correct for the extra
  level of multiple hypothesis testing.} To produce each of the
boxplots in the top row of \cref{yeast_boxplots}, we calculated the
number of statistically significant GO attributes associated with each
community, with $p=0.01$.

To measure the specificity of each method, one needs some way of
defining a false positive. One could argue that if a community is
associated with zero or very few GO attributes, then it is more likely
to be a false positive. Along these lines, we can make a couple of
observations regarding sensitivity. Seeding with the random technique
led to the detection of approximately twice as many communities as the
other seeding techniques. We observe that although the random seeding
had good sensitivity with regard to the ground truth, around half of
the communities it found had no significant GO associations, meaning
they are likely false positives. We can also observe that around 75\%
of the communities found by seeding with the clique-based seeding
strategies (DistinctC and 3-clique-perc) had 7 or more significant GO
associations (which corresponds to the green line). Because these
methods produced relatively few communities with zero or very few GO
associations, it can be argued that in this experiment they were the
most specific seeding techniques.

Comparing the results of these empirical benchmarks to the results of
the synthetic benchmarks reveals some interesting similarities and
differences. The distinct clique seeding strategy, especially when
used with either GCE or local modularity as the expansion technique,
is the only seeding strategy that performs well both in the synthetic
and empirical benchmarks. The clique percolation method performed well
on the synthetic data, but performed especially poorly on the Caltech
data set, whereas RaRe did well in the empirical benchmarks but
performed particularly poorly in the synthetic benchmarks. One cause
for the difference may be that while in the synthetic networks, each
node belongs to up to five communities, in the empirical datasets used
here, the community structure may not overlap so heavily.

\section{Discussion}
\label{discussion}
\subsection{Poor recall of Link Aggregate, RaRe, RaReCore, and Random}
In the synthetic benchmarks the poor performing seeding schemes (Link
Aggregate, RaRe, RaReCore, and Random) failed because of low
sensitivity. As each of the five successive graphs contain more and
more communities, these schemes produce fewer, rather than more,
seeds.

RaRe works on the assumption that the high-ranking nodes (i.e., hubs)
tend to be special in that they span multiple communities, while low
ranking nodes belong to only one community. If that is true, then when
one removes a relatively small proportion of high-ranking nodes, the
graph should break up into many components, each of which corresponds
to some community. However, if each node belongs to many overlapping
communities (even the low ranking nodes), then so many nodes will need
to be removed to break the network up into small disconnected
components that few cores will remain. The small number of cores
causes this seeding strategy have low sensitivity.

For a community to be seeded, the Link Aggregate method requires that
at least one of that community's nodes does not yet belong to any
other community's seed when it is iterated over. This requirement may,
in graphs with pervasive overlap, lead to many unseeded communities.
However, this method did not even work on synthetic graphs with
disjoint community structure, pointing to more fundamental problems.

The random seeding strategy has a basic arbitrariness that causes
trouble in graphs with pervasive overlap. Remember, the stopping
criterion for selecting new random seeds is that each node must belong
to at least one community. However, it would be just as valid (and
arbitrary) if one seeded the graph until each node belongs to three
communities. In graphs in which every node belongs to several
communities, leaving this value set to one has the consequence that
the random seeding prematurely terminates.

In general, RaRe, RaReCore, Link Aggregate, and the random seeding
technique all place an upper bound on the number of seeds that can be
detected, namely, $|V|$. This upper bound could be problematic for
networks with pervasive overlap, in which the number of communities
can easily be much larger. Even if there are not more communities than
nodes, if certain dense areas of the graph contain a relatively high
number of communities, then this dense region may form a subgraph with
more communities than nodes. These regions will not be sufficiently
seeded by these methods.

We note in passing that in some networks, there might be many nodes
that belong to no community whatsoever. In such cases, by forcing all
nodes into communities, these methods (except for RaReCore) may have
low specificity because they may detect seeds in regions where no
communities exist. The yeast interactome, for example, contains many
low degree nodes that may not belong to any community. Nevertheless,
the random technique forces every node into at least one community,
which may have led to the poor specificity observed on that network.

\subsection{Critical threshold of clique percolation}
Although the clique percolation method for detecting seeds worked well
on the synthetic graphs, it performed especially poorly on the Caltech
data.  In the Caltech network, the parameter value of $k=4$ is low
enough to above the critical value, at which the giant k-clique
community emerges. This explains why only three seeds were detected,
the first containing 673 nodes (or 88\% of the graph), while the other
two each contained four nodes. Similar results are obtained when $k$
takes the value 5-9---for these values, over half of the nodes in the
network are contained in the largest seed. Using values of $k$ larger
than 9 is problematic because it is not reasonable to assume that
every community which we want to find contains at least one 10-clique.
Thus no parameter value leads to acceptable results.

The results were better on the yeast interactome, partially because
the two empirical networks have quite different clique densities. The
Caltech network contains 31745 maximal cliques with four or more nodes
(41.3/node), while the yeast interactome contains 4485 such clique
(2.8/node). However, because we lowered the value of $k$ to three for
the yeast interactome to improve clique percolation's ability to pick
up the small structures that are common for protein complex, there
existed one seed that was much larger than all the others, containing
358 proteins. It seems that for a certain dense part of the graph,
cliques percolated into a seed larger than desired.

The synthetic data is generated by embedding denser $G(N,P)$ graphs
into a less dense $G(N,P)$ graph. Because this process (rather
unrealistically) does not include any mechanism that leads to a high
clustering coefficient, cliques in these graphs tend to be rare and
not adjacent to each other when compared with empirical data. Thus, on
these graphs the clique percolation method of seeding did not have the
problem of no ideal value for $k$.

\subsection{Advantages of distinct cliques as seeds}
Seeding with distinct cliques leads to higher specificity and
sensitivity because while such cliques are rare in the graph in
general, they are common in communities. The strategy should work well
if one can select a value for $k$ such that: (1) it is likely that any
clique of size $k$ is embedded in a community, and (2) nearly all
communities that ought to be found contain at least one clique of size
$k$ or larger.

While cliques are fundamental to both clique percolation and the
distinct seeding strategy, the latter does not perform any percolation
and so it does not suffer from the critical value problem.

\section{Conclusion \& Future Work}
Our benchmarks indicate that seeding methods which work well for
detecting disjoint or nearly disjoint community structure may perform
poorly on networks in which communities pervasively overlap. Many of
the methods proposed for detecting overlapping community structure
rely on the implicit assumption that most nodes belong to one
community---such methods put an arbitrary upper bound of $|V|$ on the
number of possible seeds. This assumption is inappropriate for
networks with pervasive overlap, and as a result these perform poorly
in such networks.

We considered two clique-based approaches for detecting seeds that do
not make such an assumption: distinct cliques and clique
percolation. Clique percolation is problematic because if the value
for $k$ is too low, then this method will detect one giant seed,
whereas if it is too high, then any communities lacking cliques of
size $k$ will not be seeded.  Seeding with distinct cliques worked
well in all of the benchmarks that we considered.

We have focused on seeding, which is just one aspect of local
community detection. It is still unclear which local fitness function,
expansion strategy, and termination criterion should be used. The
seeding strategies described here may also be used in heuristics that
optimize a global fitness function.

A more pressing matter, however, is benchmarking these
algorithms. While \citeauthor{Ahn2010} made good progress on this
topic in order to demonstrate that their community finding algorithm
works on empirical graphs, their benchmarking data and software is not
publicly available. It is our opinion that before future work can make
great progress on any of these topics, better empirical benchmarking
techniques must be developed and made widely available.

\section{Acknowledgments}
This work is supported by Science Foundation Ireland under grant
08/SRC/I1407, Clique: Graph and Network Analysis Cluster. We thank
Colm Ryan for his contribution to the experiments based on the yeast
interactome.

%

\end{document}